\documentclass[11pt, reqno,preprint]{article}
\usepackage{jheppub}
\usepackage{epsfig}
\usepackage{amssymb}
\usepackage{amsmath}
\usepackage{mathrsfs}
\usepackage{hyperref}
\usepackage{multirow}

\def\be{\begin{equation}}
\def\ee{\end{equation}}
\def\ba{\begin{eqnarray}}
\def\ea{\end{eqnarray}}
\newcommand{\bea}{\begin{eqnarray}}
\newcommand{\eea}{\end{eqnarray}}

\def\to{\rightarrow}

\def\e{\epsilon}

\def\sect#1{section~{\ref{#1}}}
\def\eqn#1{eq.~(\ref{#1})}

\def\eqns#1#2{eqs.~(\ref{#1}) and~(\ref{#2})}

\def\eqn#1{eq.~(\ref{#1})}
\def\eqns#1#2{eqs.~(\ref{#1}) and~(\ref{#2})}

\title{The Principle of Maximal Transcendentality
and the Four-Loop Collinear Anomalous Dimension}

\author{Lance~J.~Dixon}

\affiliation{SLAC National Accelerator Laboratory,
Stanford University, Stanford, CA 94309, USA}\emailAdd{lance@slac.stanford.edu}

\abstract{We use the principle of maximal transcendentality and
  the universal nature of subleading infrared poles to
  extract the analytic value of the four-loop collinear anomalous dimension
  in planar ${\cal N}=4$ super-Yang-Mills theory from recent QCD results,
  obtaining $\hat{\cal G}_{0}^{(4)} =
    - 300 \zeta_7 - 256 \zeta_2 \zeta_5 - 384 \zeta_3 \zeta_4$.
  This value agrees with a previous numerical result to within $0.2\%$.
  It also provides the Regge trajectory, threshold soft anomalous dimension
  and rapidity anomalous dimension through four loops.}

\dedicated{Dedicated to the memory of Lev Lipatov}

\begin{document}

\begin{flushright} SLAC--PUB--17191 \end{flushright}

\maketitle

\section{Introduction}

In the modern approach to scattering amplitudes, ${\cal N}=4$ super Yang-Mills
(SYM) theory~\cite{Brink1976bc,Gliozzi1976qd} has played a key conceptual role,
especially in the planar limit of a large number of colors where the theory
becomes integrable~\cite{Minahan2002ve,Beisert2003tq,Beisert2006ez,%
Beisert2010jr,Basso2013vsa,Gromov2013pga}
and its amplitudes become dual to Wilson loops~\cite{Alday2007hr,%
Drummond2007aua,Brandhuber2007yx,Drummond2007cf,Drummond2007au,
Bern2008ap,Drummond2008aq}.
So much is known about the analytic structure of scattering
amplitudes in planar ${\cal N}=4$ SYM that the first amplitude
with non-trivial kinematic dependence, the six-point amplitude,
can be bootstrapped to at least five loops~\cite{DelDuca2009au,%
DelDuca2010zg,Goncharov2010jf,Dixon2011pw,Dixon2013eka,Dixon2014voa,
CaronHuot2016owq}.

Ironically, the finite parts of the six-point amplitude (the remainder
function and ratio function), which are polylogarithmic
functions of three variables,
are now known to higher loop orders than is the 
dimensionally-regulated infrared-divergent prefactor
--- the BDS ansatz~\cite{Bern2005iz} --- even though the latter depends
only on four constants per loop order.  One of these constants,
the (light-like) cusp anomalous
dimension~\cite{Korchemsky1987wg,Korchemsky1991zp}
is known to all orders in planar ${\cal N}=4$ SYM,
thanks to integrability~\cite{Beisert2006ez}.
The cusp anomalous dimension controls the double pole in $\e$
in the logarithm of the dimensionally regularized BDS ansatz.
The single pole is controlled by the ``collinear'' anomalous dimension.
In planar ${\cal N}=4$ SYM, it is known analytically through
three loops~\cite{Bern2005iz,Moch2005id}, and it was computed
numerically at four loops a decade ago~\cite{Cachazo2007ad}.
The nonplanar contribution to the four-loop collinear anomalous
dimension was computed numerically very
recently~\cite{Boels2017ftb,Boels2017fng}.
The collinear anomalous dimension also enters the Regge
trajectory for forward
scattering~\cite{Drummond2007aua,Naculich2007ub,DelDuca2008pj}.
An eikonal (Wilson line) version of it enters both
the threshold soft anomalous dimension for threshold
resummation~\cite{Moch2005tm,Becher2007ty,Li2014afw,Li2016ctv}
and the rapidity anomalous dimension for transverse momentum
resummation~\cite{Chiu2012ir,Li2016axz,Li2016ctv,Vladimirov2016dll}.

The BDS ansatz also contains two finite constants at each loop order,
one for the four-point amplitude and one for the five-point amplitude.
One of these constants is known analytically at three loops~\cite{Bern2005iz},
but the other is only known numerically at this loop
order~\cite{Spradlin2008uu}.

The purpose of this paper is to provide an analytical value
for one of the four constants in question at four loops, namely
the collinear anomalous dimension in planar ${\cal N}=4$ SYM.
We do so
by leveraging two recent four-loop computations in QCD in the large $N_c$
limit~\cite{Lee2016ixa,Moch2017uml},
as well as the principle of maximal
transcendentality~\cite{Kotikov2001sc,Kotikov2002ab,Kotikov2004er,%
Kotikov2007cy}.
This principle states that for suitable quantities, such as the BFKL
and DGLAP kernels, the result in ${\cal N}=4$ SYM can be obtained
from that in QCD by converting the fermion representation from the
fundamental (quarks) to the adjoint (gluinos) and then keeping only
the functions that have the highest transcendental weight.  In momentum
space ($x$-space) these functions are typically iterated integrals, and
the weight is the number of iterations; in Mellin-moment space,
it corresponds to the number of sums in the nested sums~\cite{Remiddi1999ew}.
Here we will only need the notion of weight for ordinary Riemann zeta values,
$\zeta_n \equiv \zeta(n)$, for which the weight is $n$. Also, the weight is
additive for products, and rational numbers have weight zero.

The complete set of observables to which this principle can be applied
is still unclear.
Besides anomalous dimensions, it has also been successfully applied to
form factors, matrix elements of gauge-invariant operators with two
or three external partons~\cite{%
Brandhuber2012vm,Loebbert2015ova,Brandhuber2016fni,Loebbert2016xkw,
Brandhuber2017bkg},
and to certain configurations of semi-infinite Wilson
lines~\cite{Li2014afw,Li2016ctv}.
However, it does not hold for scattering
amplitudes with four or five external
gluons, even at one loop~\cite{Bern1993mq}, in the sense that there
are maximally transcendental parts of the QCD one-loop amplitudes
which have different rational prefactors from the corresponding
${\cal N}=4$ SYM amplitudes.

Here we will apply the maximal transcendentality principle
to the collinear anomalous dimension.
This quantity depends on the method of regularization.  We will compute
its value in dimensional regularization
--- in fact, in a supersymmetric version
of dimensional regularization such as dimensional reduction.
The collinear anomalous dimension
has also been computed using the so-called massive, or Higgs,
regularization~\cite{Alday2009zm,%
Henn2010bk,Henn2010ir,Drummond2010mb,Henn2011by}.
The Higgs-regulated result begins to differ from the
dimensionally-regularized value starting at three loops~\cite{Henn2011by},
the last value for which it is known.
A dual conformal regulator for infrared divergences has also
been defined~\cite{Bourjaily2013mma}; however, the multi-loop values
of the collinear anomalous dimension in this scheme are still
under investigation~\cite{BDDP}.

One might think that the collinear anomalous dimension
in planar ${\cal N}=4$ SYM could simply be read off from the leading
transcendental terms in the large-$N_c$
quark collinear anomalous dimension~\cite{Lee2016ixa}.
However, the full-color expression for this quantity is a polynomial
in the adjoint and fundamental quadratic Casimirs, $C_A$ and $C_F$.
In the large-$N_c$ limit, $C_A \to N_c$ while $C_F \to N_c/2$.
In order to apply the principle of maximal transcendentality
at large $N_c$, we should first set $C_F \to C_A$; that is,
$C_A \to N_c$ and $C_F \to C_A \to N_c$, not $N_c/2$.  There is not enough
information left in the large-$N_c$ limit to make the correct replacement.

However, if we can first convert the collinear anomalous dimension
to an appropriate eikonal (Wilson line) quantity, then we can make the
correct replacement.  In a conformal theory, this ``eikonal bypass'' only
requires~\cite{Dixon2008gr} knowledge of the virtual anomalous dimension,
the coefficient of $\delta(1-x)$ in the DGLAP kernel.
The virtual anomalous dimension in large-$N_c$ QCD was computed
recently at four loops~\cite{Moch2017uml}, and we can make use
of its leading transcendental part to do the conversion.
Once we have the eikonal quantity, we use its non-abelian
exponentiation property~\cite{Gatheral1983cz,Frenkel1984pz},
which means that it is ``maximally non-abelian''.  That is, at any loop order,
it can contain only one quadratic Casimir for the representation of the
Wilson line; the remaining group theory factors must all be $C_A$.
(A subtlety that arises when quadratic Casimir scaling does not hold
is addressed in \sect{ComputationSection}.)
This information suffices to allow us to apply the principle of maximal
transcendentality and extract the eikonal quantity in planar ${\cal N}=4$ SYM.
Then we use the virtual anomalous dimension in planar ${\cal N}=4$ SYM,
which has been computed to all orders using
integrability~\cite{Freyhult2007pz,Freyhult2009my,Fioravanti2009xt},
to convert back to the non-eikonal collinear anomalous dimension.

This paper is organized as follows.  In \sect{ReviewSection} we briefly
review the infrared structure of scattering amplitudes, form factors and
Wilson loops in planar ${\cal N}=4$ SYM.  In \sect{ComputationSection}
we carry out the computation and then conclude.


\section{Review of Infrared Structure of Planar 
\texorpdfstring{${\cal N}=4$}{N=4} SYM}
\label{ReviewSection}

In this section we give a very brief review of the infrared structure
of scattering amplitudes, form factors and Wilson loops
in planar ${\cal N}=4$ SYM.
In general, multi-loop $n$-point amplitudes can be factorized into soft,
collinear and hard virtual contributions, where soft gluon exchange
can connect any of the $n$ hard external
legs~\cite{Kidonakis1998bk,Kidonakis1998nf}.
This factorization has consequences for the infrared poles in $\e$
of dimensionally-regularized multi-loop
amplitudes~\cite{Catani1998bh,Sterman2002qn}.

In the planar limit, the soft structure simplifies enormously, because
only color-adjacent lines can exchange soft gluons, and the infrared
structure of the amplitude for $n$ external adjoint particles becomes
the product of $n$ ``wedges'', each equivalent to the square root of
a Sudakov form factor for producing two adjoint particles~\cite{Bern2005iz}.
The infrared behavior of the Sudakov form factor was studied using
factorization and renormalization group evolution, beginning in 
the 1970s~\cite{Mueller1979ih,Collins1980ih,Sen1981sd,
Collins1989bt,Korchemsky1988pn,Magnea1990zb}.
Besides the $\beta$ function (which
of course vanishes in ${\cal N}=4$ SYM), the only quantities
that enter are the (light-like) cusp anomalous dimension
$\gamma_K$~\cite{Korchemsky1987wg,Korchemsky1991zp}
and an integration constant for a function ${\cal G}(q^2)$,
which we will refer to as the collinear anomalous dimension
and denote by ${\cal G}_0$~\cite{Magnea1990zb,Sterman2002qn}.

We consider gauge group $SU(N_c)$ and adopt the ``integrability'' notation
for the large-$N_c$ coupling constant,
\be 
g^2\ \equiv\ N_c\frac{g_{\rm YM}^2}{(4\pi)^2}\ =\ C_A \frac{\alpha_s}{4\pi}
\ =\ \frac{\lambda}{(4\pi)^2}\ =\ \frac{a}{2} \,,
\label{gsqdef}
\ee
where $\alpha_s=g_{\rm YM}^2/(4\pi)$, $\lambda = N_c g_{\rm YM}^2$
is the 't~Hooft coupling, and $a$ was used e.g.~in ref.~\cite{Bern2005iz}.
The quadratic Casimir in the adjoint representation is $C_A  = N_c$,
while in the fundamental representation it is $C_F  = (N_c^2-1)/(2N_c)$.

We expand the cusp and collinear anomalous dimensions in terms of $g^2$:
\bea
\gamma_K(g) &=& \sum_{L=1}^\infty g^{2L} \hat\gamma_K^{(L)} \,, \label{gammaKexp}\\
{\cal G}_0(g) &=& \sum_{L=1}^\infty g^{2L} \hat{\cal G}_0^{(L)} \,.
 \label{G0xp}
\eea
(Note that another normalization is often used for the cusp anomalous
dimension, $\gamma_K = 2\Gamma_{\rm cusp}$.)
The cusp anomalous dimension is known to all orders, thanks to 
integrability~\cite{Beisert2006ez}.  The first four terms in its
perturbative expansion are:
\be
\gamma_K^{\rm planar\ {\cal N}=4}\ =\ 8 \, g^2 - 16 \, \zeta_2 \, g^4
+ 176 \, \zeta_4 \, g^6
- \Bigl( 1752 \, \zeta_6 + 64 \, (\zeta_3)^2 \Bigr) g^8 \,.~~
\label{cuspplanarNeq4}
\ee
We give the previously-known three-loop result for ${\cal G}_0(g)$ below,
in \eqn{G0_Neq4}.

In a non-conformal theory, when the differential equation
for the Sudakov form factor is integrated up, infrared poles are obtained
that involve integrals over functions of the running coupling in
$D=4-2\e$ dimensions.  Because planar ${\cal N}=4$ SYM is conformally
invariant, the integrals can be performed analytically.
One obtains for the color-ordered $n$-point scattering amplitude
$A_n$~\cite{Bern2005iz}:
\be
\ln\biggl( \frac{A_n}{A_n^{\rm tree}} \biggr) 
= -\frac{1}{8} \sum_{L=1}^\infty \frac{g^{2L}}{L^2 \, \e^2}
  \Bigl( \hat\gamma_K^{(L)} + 2L \, \e \, \hat{\cal G}_0^{(L)} \Bigr)
  \sum_{i=1}^n \biggl( \frac{\mu^2}{-s_{i,i+1}} \biggr)^{L\e}\ +\ \hbox{finite},
\label{npointIRpoles}
\ee
where $s_{i,i+1} = (k_i+k_{i+1})^2$.

The form factor $F(Q^2)$ for producing two adjoint particles corresponds
to setting $n=2$ in this formula, in which case the two wedges
have the same kinematics,
\be
\ln F(Q^2)
= -\frac{1}{4} \sum_{L=1}^\infty \frac{g^{2L}}{L^2 \, \e^2}
  \Bigl( \hat\gamma_K^{(L)} + 2L \, \e \, \hat{\cal G}_0^{(L)} \Bigr)
  \biggl( \frac{\mu^2}{-Q^2} \biggr)^{L\e} \ +\ \hbox{finite},
\label{FFIRpoles}
\ee

Wilson loops for light-like $n$-gons $C_n$ contain ultraviolet poles
rather than infrared ones.  These poles
have a very similar form (with $\e \to -\e$ due to their
ultraviolet nature)~\cite{Drummond2007aua}:
\be
\ln W_{C_n} = -\frac{1}{8} \sum_{L=1}^\infty \frac{g^{2L}}{L^2 \, \e^2}
  \Bigl( \hat\gamma_K^{(L)} - 2L \, \e \, \hat{\cal G}_{0,\,{\rm eik}}^{(L)} \Bigr)
  \sum_{i=1}^n \biggl( \frac{\mu_{\rm UV}^2}{-x_{i,i+2}^2} \biggr)^{-L\e}
  \ +\ \hbox{finite},
\label{WilsonLoopUVpoles}
\ee
where $x_{i,i+2}^2 = (x_i-x_{i+2})^2$ are invariant distances between
the corners of the polygons $x_i^\mu$.  The amplitude-Wilson loop duality
makes the identification $(x_i-x_{i+2})^2 = (k_i+k_{i+1})^2$. 
While the leading double poles in Wilson loops
are governed by the same quantity as in amplitudes, namely
$\gamma_K$, a different quantity appears in the subleading poles,
${\cal G}_{0,\,{\rm eik}}$, whose expansion is defined by
\be
{\cal G}_{0,\,{\rm eik}}(g)
= \sum_{L=1}^\infty g^{2L} \hat{\cal G}_{0,\,{\rm eik}}^{(L)} \,.
\label{G0eikxp}
\ee
instead of ${\cal G}_{0}$.

The relation between ${\cal G}_0$ and ${\cal G}_{0,\,{\rm eik}}$ was
explored in ref.~\cite{Dixon2008gr}, where it was shown that for
a conformal theory, they obey a particularly simple relation,
\be
{\cal G}_0 = {\cal G}_{0,\, {\rm eik}} + 2 \, B \,.
\label{GGeikBrelation}
\ee
(Empirical evidence for this kind of relation was given in
refs.~\cite{Ravindran2004mb,Moch2005tm}.)
Here $B$, sometimes called $B_\delta$ or the virtual anomalous dimension,
is the coefficient of the first subleading term in the limit as $x\to1$
of the DGLAP kernel for parton $i$ to split to parton $i$:
\be
P_{ii}(x)\ =\ \frac{\gamma_K}{2(1-x)_+} + B_i \, \delta(1-x) + \ldots.
\label{Piilargex}
\ee
In a general theory, $B=B_i$ depends on the type of parton $i$
(also the leading, cusp, term in \eqn{Piilargex} depends on the color
representation of parton $i$),
but in ${\cal N}=4$ SYM $B$ is the same for all partons, by supersymmetry.

In planar ${\cal N}=4$ SYM, thanks to dual conformal symmetry,
the gluon Regge trajectory governing the forward limit of the four-point
amplitude can be computed from the cusp and collinear
anomalous dimensions~\cite{Drummond2007aua,Naculich2007ub,DelDuca2008pj}.
The result is~\cite{Drummond2007aua}
\be
\frac{\partial}{\partial\ln s} \ln A_4(s,t) \Big|_{s\gg t}\ =\ \omega_R(-t),
\label{omegaRdef}
\ee
where
\be
\omega_R(-t)\ =\ \frac{1}{4} \gamma_K \ln\biggl(\frac{\mu^2}{-t}\biggr)
+ \frac{1}{4\e} \int_0^{g^2} \frac{dg^{\prime2}}{g^{\prime2}} \gamma_K(g^\prime)
+ \frac{1}{2} {\cal G}_0 + O(\e).
\label{omegaR}
\ee
Hence our four-loop result for ${\cal G}_0$ will also provide $\omega_R(-t)$
to the same order.


\section{The Computation}
\label{ComputationSection}

In ref.~\cite{Lee2016ixa}, the quark form factor was computed to four loops
in the large $N_c$ limit of QCD, and the cusp and collinear
anomalous dimensions for large-$N_c$ QCD were determined from it.
In ref.~\cite{Kotikov2004er} it was 
proposed that the ${\cal N}=4$ super-Yang-Mills results for the
twist-two anomalous
dimensions (which includes the cusp anomalous dimension, but not
the collinear anomalous dimension) could be extracted from the
leading transcendental terms in the QCD result by setting $C_F \to C_A$.
Through three loops, where full-color QCD results are known,
the same extraction procedure also works for
the collinear anomalous dimension.

Unfortunately, as mentioned in the introduction,
the large $N_c$ limit corresponds to
\be
C_F = \frac{N_c^2-1}{2N_c} \to \frac{N_c}{2} = \frac{C_A}{2} \,.
\label{CACFlargeNc}
\ee
The factor of $1/2$ means that the $C_F \to C_A$ replacement can't be deduced
in general from the large $N_c$ limit.
However, there is a workaround, the eikonal bypass discussed
in the introduction, which involves
converting the non-eikonal quark collinear anomalous dimension
to an eikonal (Wilson line) quantity~\cite{Dixon2008gr},
with the help of the recent four-loop result for the DGLAP kernels
in the large $N_c$ limit of QCD~\cite{Moch2017uml}.
In particular, we need the coefficient of $\delta(1-x)$ in this
result, the virtual anomalous dimension.
We will see that th $C_F \to C_A$ replacement can be performed for the
eikonal quantity we have constructed.  Afterwards,
one can use the virtual anomalous dimension for planar ${\cal N}=4$
SYM~\cite{Freyhult2007pz,Freyhult2009my,Fioravanti2009xt}
to convert back to the non-eikonal collinear anomalous
dimension.  We will find an analytic expression that is
quite close to the numerical result~\cite{Cachazo2007ad}.

Through four loops, the leading transcendental part of the leading-color
quark collinear anomalous dimension
is~\cite{Lee2016ixa,Gehrmann2010ue,Henn2016men}
\be
\gamma_q|_{\rm L.C.L.T.}\ =\ 7 \, \zeta_3 \, g^4
 - \Bigl( 68 \, \zeta_5 + \frac{44}{3} \, \zeta_2 \zeta_3 \Bigr) g^6
 + \Bigl( 705 \, \zeta_7 + 144 \, \zeta_2 \, \zeta_5
        + 164 \, \zeta_3 \, \zeta_4 \Bigr) g^8 \,.
\label{gamma_q_LCLT}
\ee
Through three loops, we also know the full group-theoretical
decomposition~\cite{Gehrmann2010ue}:
\bea
\gamma_q|_{\rm L.T.} &=& C_F \Biggl\{ 
( 26 \, C_A - 24 \, C_F ) \zeta_3 \, \Biggl(\frac{\alpha_s}{4\pi}\Biggr)^2
\nonumber\\&&\null\hskip0.3cm
 - \biggl[ \Bigl( 136 C_A^2 + 120 C_F C_A - 240 C_F^2 \Bigr) \zeta_5
   + \Bigl( \frac{88}{3} C_A^2 + 16 C_F C_A - 32 C_F^2  \Bigr)
         \zeta_2 \zeta_3
         \biggr]
 \Biggl(\frac{\alpha_s}{4\pi}\Biggr)^3
\nonumber\\&&\null\hskip0.3cm
 + \ldots \Biggr\} \,.
\label{gamma_q_LT}
\eea
Letting $C_F \to C_A$, the ${\cal N}=4$ SYM result, for a gluon or
gluino in the adjoint representation is~\cite{Kotikov2003fb,Kotikov2004er}:
\be
\gamma^{{\cal N}=4} = 2 \, \zeta_3 \, \Biggl(\frac{C_A\,\alpha_s}{4\pi}\Biggr)^2
- \Bigl( 16 \, \zeta_5 + \frac{40}{3} \, \zeta_2 \zeta_3 \Bigr)
        \Biggl(\frac{C_A\,\alpha_s}{4\pi}\Biggr)^3
+ \ldots \,.
\label{gamma_Neq4}
\ee

In refs.~\cite{Sterman2002qn,Bern2005iz}, the collinear anomalous dimension
${\cal G}_0$ was evaluated to two and three loops in planar ${\cal N}=4$ SYM,
although to this order there are no subleading color terms.
The result is
\be
{\cal G}_0^{{\cal N}=4} = 
 - 4 \, \zeta_3 \, \Biggl(\frac{C_A\,\alpha_s}{4\pi}\Biggr)^2
 + \Bigl( 32 \, \zeta_5 + \frac{80}{3} \, \zeta_2 \zeta_3 \Bigr)
    \Biggl(\frac{C_A\,\alpha_s}{4\pi}\Biggr)^3
 - \ldots \,.
\label{G0_Neq4}
\ee
Comparing with \eqn{gamma_Neq4}, there is a difference in 
normalization convention by a factor of two:
${\cal G}_0 = - 2 \gamma$.

In ref.~\cite{Moch2017uml}, the twist-two anomalous dimensions
or DGLAP kernels were computed in the large $N_c$ limit of QCD to four loops.
In the limit that $x\to1$, as in \eqn{Piilargex},
the coefficient of the leading $1/(1-x)_+$ term
is the cusp anomalous dimension~\cite{Korchemsky1991zp}.
The next-to-leading term as $x\to1$ is the coefficient of $\delta(1-x)$,
sometimes called the virtual anomalous dimension, or $B_\delta$, or just $B$.
The large-$N_c$, leading transcendentality terms in $B$ for quarks are given
by~\cite{Moch2017uml,Moch2004pa}:
\be
B_q|_{\rm L.C.L.T.}\ =\ 20 \, \zeta_5 \, g^6
- \Bigl( 280 \, \zeta_7 + 40 \, \zeta_2 \, \zeta_5
        - 16 \, \zeta_3 \, \zeta_4 \Bigr) g^8 \,.
\label{B_LCLT}
\ee
Through three loops, we also know the full group-theoretical
decomposition~\cite{Moch2004pa}:
\bea
B_q|_{\rm L.T.} &=& C_F \Biggl\{ 
- 12 ( C_A - 2 C_F ) \zeta_3 \, \Biggl(\frac{\alpha_s}{4\pi}\Biggr)^2
\nonumber\\&&\null\hskip0.3cm
+ \biggl[ \Bigl( 40 \, C_A^2 + 120 \, C_F C_A - 240 C_F^2 \Bigr) \zeta_5
   + 16 C_F ( C_A - 2 C_F ) \zeta_2 \zeta_3 \biggr]
 \Biggl(\frac{\alpha_s}{4\pi}\Biggr)^3
\nonumber\\&&\null\hskip0.3cm
 + \ldots \Biggr\} \,.
\label{B_LT}
\eea
Letting $C_F \to C_A$, the ${\cal N}=4$ SYM result, for a gluon or
gluino, is~\cite{Kotikov2004er}:
\be
B^{{\cal N}=4} = 12 \, \zeta_3 \, \Biggl(\frac{C_A\,\alpha_s}{4\pi}\Biggr)^2
- \Bigl( 80 \, \zeta_5 + 16 \, \zeta_2 \zeta_3 \Bigr)
    \Biggl(\frac{C_A\,\alpha_s}{4\pi}\Biggr)^3
+ \ldots \,.
\label{B_Neq4}
\ee

We now use \eqn{GGeikBrelation} to construct the eikonal quantity
\be
{\cal G}_{0,\, {\rm eik}} = {\cal G}_{0} - 2 \, B  = - 2\,\gamma_q - 2 \, B
\label{G0eikfromG0B}
\ee
in ${\cal N}=4$ SYM for a Wilson line in the fundamental
``$F$'' representation through three loops, using \eqns{gamma_q_LT}{B_LT}:
\be
{\cal G}_{0,\, {\rm eik},F} = C_F \Biggl\{ 
- 28 \, C_A \, \zeta_3 \, \Biggl(\frac{\alpha_s}{4\pi}\Biggr)^2
+ \biggl( 192 \, \zeta_5 + \frac{176}{3} \, \zeta_2 \zeta_3 \biggr)
C_A^2 \Biggl(\frac{\alpha_s}{4\pi}\Biggr)^3
 + \ldots \Biggr\} \,.
\label{Geik_LT}
\ee
We see that all the $C_F$ terms cancel, except for the overall one.
This result reflects non-abelian exponentiation for this type of
Wilson line~\cite{Gatheral1983cz,Frenkel1984pz}.
These results agree with the leading transcendental part of the
results for $f_L^q$ in ref.~\cite{Moch2005tm}.

The threshold soft anomalous dimension $\gamma^s$ defined in
refs.~\cite{Li2014afw,Li2016ctv}
(called $\gamma^W$ in ref.~\cite{Becher2007ty}) is the same as
${\cal G}_{0,\, {\rm eik},F}$ up to a conventional minus sign,
$\gamma^s = - {\cal G}_{0,\, {\rm eik},F}$, and \eqn{Geik_LT}
agrees with the leading transcendental part of the QCD
result in refs.~\cite{Li2014afw,Li2016ctv}.
The rapidity anomalous dimension $\gamma^r$, which enters the SCET
description of transverse momentum resummation, has also 
been computed to three loops~\cite{Li2016ctv}.  The result
agrees with the threshold soft anomalous dimension, up to terms
that are proportional to coefficients of the QCD beta function.
This result was explained in ref.~\cite{Vladimirov2016dll}
by mapping the appropriate configurations of Wilson lines
for the two computations into each other using a conformal
transformation.  Hence we will obtain the four-loop values
of both the threshold soft and rapidity anomalous dimensions
in planar ${\cal N}=4$ SYM from
\be
\gamma^{s, \, \rm planar\ {\cal N}=4} = \gamma^{r, \, \rm planar\ {\cal N}=4} 
= - {\cal G}_{0, {\rm eik}}^{\rm planar\ {\cal N}=4} \,.
\label{softrap}
\ee

In planar ${\cal N}=4$ SYM, the natural Wilson line is in the adjoint
representation, not the fundamental.
In the large $N_c$ limit, this collinear anomalous dimension can 
be obtained from \eqn{Geik_LT} simply by multiplying by an overall
factor of 2, since $C_A=2C_F$ in the large $N_c$ limit.
What about at four loops?  At this order, quadratic Casimir scaling might be
violated.  That is, inspecting the color factors of all the Feynman
diagrams that contribute at this order, we see that
${\cal G}_{0,\, {\rm eik},F}$ might contain --- besides
$C_F$ times a polynomial in $C_A$ --- a color factor of
\be
 \frac{d_F^{abcd} d_A^{abcd}}{N_F} = \frac{(N_c^2-1)(N_c^2+6)}{48} \,.
\label{dFdA}
\ee
(See e.g.~eq.~(2.14) of ref.~\cite{Herzog2017ohr}.)
If so, the corresponding term in the case of an adjoint Wilson line
would have the same numerical coefficient multiplying
\be
 \frac{d_A^{abcd} d_A^{abcd}}{N_A} = \frac{N_c^2(N_c^2+36)}{24} \,.
\label{dAdA}
\ee
However, the latter factor is precisely twice the former factor
in the large $N_c$ limit, which is the same factor as for the conversion
$C_F \to C_A$ in this limit.
Given that there are no $C_F$ terms in ${\cal G}_{0,\, {\rm eik},F}$ except
for the overall $C_F$, ${\cal G}_{0,\, {\rm eik}}$
in the large $N_c$ limit of ${\cal N}=4$ SYM can be extracted from
the leading transcendality terms of the corresponding eikonal
quantity in the large $N_c$ limit of QCD.
(The beta-function correction terms to \eqn{GGeikBrelation} for
a non-conformal theory are also subleading in transcendentality.)

In summary, the eikonal collinear anomalous dimension for
planar ${\cal N}=4$ SYM can be obtained from the large-$N_c$ QCD
results for $\gamma_q$ and $B_q$ through four loops, using
\be
{\cal G}_{0, {\rm eik}}^{\rm planar\ {\cal N}=4}
= 2 \, \Bigl( -2 \gamma_q|_{\rm L.C.L.T.} - 2 B_q|_{\rm L.C.L.T.} \Bigr) \,.
\label{G0eikplanarNeq4formula}
\ee
Inserting \eqns{gamma_q_LCLT}{B_LCLT}, we obtain
\be
{\cal G}_{0, {\rm eik}}^{\rm planar\ {\cal N}=4}\ =\ - 28 \, \zeta_3 \, g^4
+ \Bigl( 192 \, \zeta_5 + \frac{176}{3} \, \zeta_2\zeta_3 \Bigr) \, g^6
- \Bigl( 1700 \, \zeta_7 + 416 \, \zeta_2 \, \zeta_5
        + 720 \, \zeta_3 \, \zeta_4 \Bigr) g^8 \,.~~~~~~
\label{G0_eik_planarNeq4}
\ee
The virtual anomalous dimension in planar ${\cal N}=4$ SYM
is known to all orders from
integrability~\cite{Freyhult2007pz,Freyhult2009my,Fioravanti2009xt}:
\bea
B^{\rm planar\ {\cal N}=4}\ =\ 12 \, \zeta_3 \, g^4
- \Bigl( 80 \, \zeta_5 + 16 \, \zeta_2\zeta_3 \Bigr) \, g^6
+ \Bigl( 700 \, \zeta_7 + 80 \, \zeta_2 \, \zeta_5
        + 168 \, \zeta_3 \, \zeta_4 \Bigr) g^8 + \ldots \,.~~~~
\label{B_planarNeq4}
\eea
We set $L=2$ in eq.~(3.16) of ref.~\cite{Freyhult2007pz}, and multiply
by $-1/2$ to account for the different normalization convention.

The non-eikonal collinear anomalous dimension in planar ${\cal N}=4$ SYM
is then:
\bea
{\cal G}_{0}^{\rm planar\ {\cal N}=4} &=&
{\cal G}_{0, {\rm eik}}^{\rm planar\ {\cal N}=4} + 2 B^{\rm planar\ {\cal N}=4}
\nonumber\\
&=&
- 4 \, \zeta_3 \, g^4
+ \Bigl( 32 \, \zeta_5 + \frac{80}{3} \, \zeta_2\zeta_3 \Bigr) \, g^6
- \Bigl( 300 \, \zeta_7 + 256 \, \zeta_2 \, \zeta_5
        + 384 \, \zeta_3 \, \zeta_4 \Bigr) g^8 \,. \nonumber\\
&~& \label{G0_planarNeq4}
\eea
The numerical value of the four-loop coefficient is
\be
-1238.7477172547735332918988\ldots
\label{G04loopnum}
\ee
which can be compared with the number from ref.~\cite{Cachazo2007ad}:
\be
-1240.9(3).
\label{G04loopCSVnum}
\ee
The two results are within about $0.2\%$, although they are not
within the error budget of $0.3$ reported in ref.~\cite{Cachazo2007ad}.
It would be very nice to check the analysis in this paper with an improved
numerical value.

The first order at which ${\cal G}_{0}^{{\cal N}=4}$ can have a subleading-color
term is four loops.  Recently this term
has been computed numerically~\cite{Boels2017ftb,Boels2017fng},
\be
{\cal G}_{0,\,{\rm NP}}^{(4),\,{\cal N}=4}
 = - 384 \times ( - 17.98 \pm 3.25 ) \, \frac{1}{N_c^2} \,.
\label{subleadingcolornumeric}
\ee
Could one try to get an analytic value for this quantity
using the methods in this paper?

One issue is that the principle of maximal transcendentality has not really
been tested yet for cases where there is a subleading-color contribution
to ${\cal N}=4$ SYM, but one could try nevertheless.
The good news is that the simple relation~(\ref{GGeikBrelation}) continues
to hold at subleading color
--- whereas in a non-conformal theory it would
receive additional corrections
depending on the infrared-finite part of a form factor~\cite{Dixon2008gr}.
The bad news is that there are not yet analytic values for the
subleading-color terms in QCD at four loops, for either $\gamma_q$ or $B_q$. 
(Approximate numerical values are available for
$B_q$~\cite{Moch2017uml}.)
Once they become available, it will be possible to compute
\eqn{subleadingcolornumeric} analytically, if it is not already known
by then.  In fact, the eikonal bypass of using \eqn{GGeikBrelation} should
become unnecessary at that point, once the full color dependence
of the QCD result for $\gamma_q$ is known.

In summary, in this paper we obtained an analytical value~(\ref{G0_planarNeq4})
for the four-loop collinear anomalous dimension in planar ${\cal N}=4$ SYM,
which also provides the Regge trajectory, threshold soft anomalous dimension
and rapidity anomalous dimension at this order.
We hope that this additional data point will inspire those
versed in integrability methods to try to compute this quantity to all
loop orders!

\vskip0.5cm

\noindent {\bf Acknowledgments}

I am grateful to Benjamin Basso, Tomasz Lukowski, Mark Spradlin,
Matthias Staudacher and HuaXing Zhu for useful discussions,
and to HuaXing Zhu for very helpful comments on the manuscript.
This research was supported by the US Department of Energy under
contract DE--AC02--76SF00515, and by the Munich Institute for Astro-
and Particle Physics (MIAPP) of the DFG cluster of excellence
``Origin and Structure of the Universe''.


\end{document}